\documentclass[a4paper,11pt]{article}
\pdfoutput=1

\usepackage{soul}
\usepackage{jcappub} 
\usepackage{tikz}
\usepackage{xcolor}

\usepackage[utf8]{inputenc}
\usepackage{amsmath, amsthm, mathtools}
\usepackage{braket}

\usepackage{tensor}
\usepackage{verbatim}
\usepackage{amsmath,amsfonts,amssymb}

\usepackage{fontawesome}

\usepackage{graphicx}
\usepackage{adjustbox}
\usepackage{tabularx}
\usepackage{tensor}
\usepackage{amsbsy}
\usepackage{braket}
\usepackage{slashed}
\usepackage{enumitem}
\hypersetup{%
,urlcolor=blue
,citecolor=blue
,linkcolor=blue
}

\usetikzlibrary{shapes.geometric, arrows}

\tikzstyle{startstop} = [rectangle, rounded corners, minimum width=3cm, minimum height=1cm,text centered, draw=black, fill=red!30]
\tikzstyle{process} = [rectangle, minimum width=3cm, minimum height=1cm, text centered, draw=black, fill=orange!30]
\tikzstyle{decision} = [diamond, minimum width=3cm, minimum height=1cm, text centered, draw=black, fill=green!30]
\tikzstyle{arrow} = [thick,->,>=stealth]

\usepackage[T1]{fontenc} 

\usepackage{braket}
\title{\boldmath Axion-Photon Mixing in 3D: Classical Equations and Geometric Optics}

\author[]{J. I. McDonald,}
\author[]{P. Millington}

\emailAdd{jamie.mcdonald@manchester.ac.uk}

\affiliation[]{
Department of Physics and Astronomy\\
University of Manchester\\
Manchester, M13 9PL, UK}

\abstract{Light particle-photon mixing in magnetised plasmas plays a vital role in constraining the existence of new physics, especially axions, dark photons, and ultra-high-frequency gravitational waves. Recently, we derived an expression for the resonant conversion of axions to photons in inhomogeneous media using kinetic theory to derive photon transport equations. In this work, we show how the same expression for the conversion probability can be obtained from the classical wave equations of axion-electrodynamics by deriving an equivalent transport equation along the photon worldline. This result provides further corroboration of this expression for the resonant production of photons from light particles, which has also recently been supported by independent numerical simulations of full axion-electrodynamics. In addition, this new approach provides a more general expression that accounts for mixing away from resonance, which is integrated along the whole worldline of the photon in a way that naturally incorporates a curved photon trajectory relevant to refractive media where the photon and light-particle worldlines differ. }

\bigskip

\tikzstyle{startstop} = [rectangle, rounded corners, minimum width=3cm, minimum height=1cm, text centered, draw=black, fill=red!30]
\tikzstyle{process} = [rectangle, minimum width=4cm, minimum height=1cm, text centered, draw=black, fill=orange!30]
\tikzstyle{stop} = [rectangle, rounded corners, minimum width=3cm, minimum height=1cm, text centered, draw=black, fill=yellow!30] 
\tikzstyle{arrow} = [thick,->,>=stealth]

\begin{document}

\notoc

\maketitle
\flushbottom

\section{Introduction}

Plasmas provide one of the most versatile environments in which to test physics beyond the Standard Model. In particular, the mixing of photons with other light particles has been the subject of intense research for many decades, and mixing of axions and gravitons  in plasmas has given rise to a wide range of probes of new physics --- see, e.g., \cite{Caputo:2024oqc,Aggarwal:2020olq} for reviews. Radio observations of neutron stars now offer some of the most powerful probes of dark matter axions \cite{Darling:2020plz,Darling:2020uyo,Battye2022,FosterSETI2022,Battye:2023oac}. This probe exploits resonant effects in which axion-photon conversion is kinematically enhanced in regions where the dispersion relation of the photon matches that of the other light particle. Starting from early beginnings \cite{Raffelt:1987im,pshirkov2009} and a subsequent revival of interest \cite{Hook:2018iia,Huang:2018lxq,Battye_2020}, the study of radio production from axion dark matter in the magnetospheres of neutron stars has since grown into a sophisticated sub-discipline, employing ray-tracing techniques \cite{Leroy:2019ghm,Witte:2021arp,Battye:2021xvt,McDonald:2023shx,Tjemsland:2023vvc} to model the transport of photons. 

There has also been significant focus on the production process of axions from photons itself, beginning with early  treatments \cite{Battye_2020}, where it was shown how to reduce axion-electrodynamics to 1D mixing equations under very restricted conditions in an isotropic plasma. For some time, systematic calculations of the resonant production of axions from photons in 3D, incorporating the full range of physical effects (refraction, multiple polarisations, and anisotropy) had remained an unsolved problem. Last year, we offered a comprehensive calculation based on kinetic theory \cite{McDonald:2023ohd}, which also provides a direct link with ray-tracing approaches (see Ref.~\cite{McDonald:2023shx}) needed to numerically model signatures for use in radio astronomy. This kinetic-theory result has been shown to have good agreement with numerical simulations~\cite{Gines:2024ekm}.

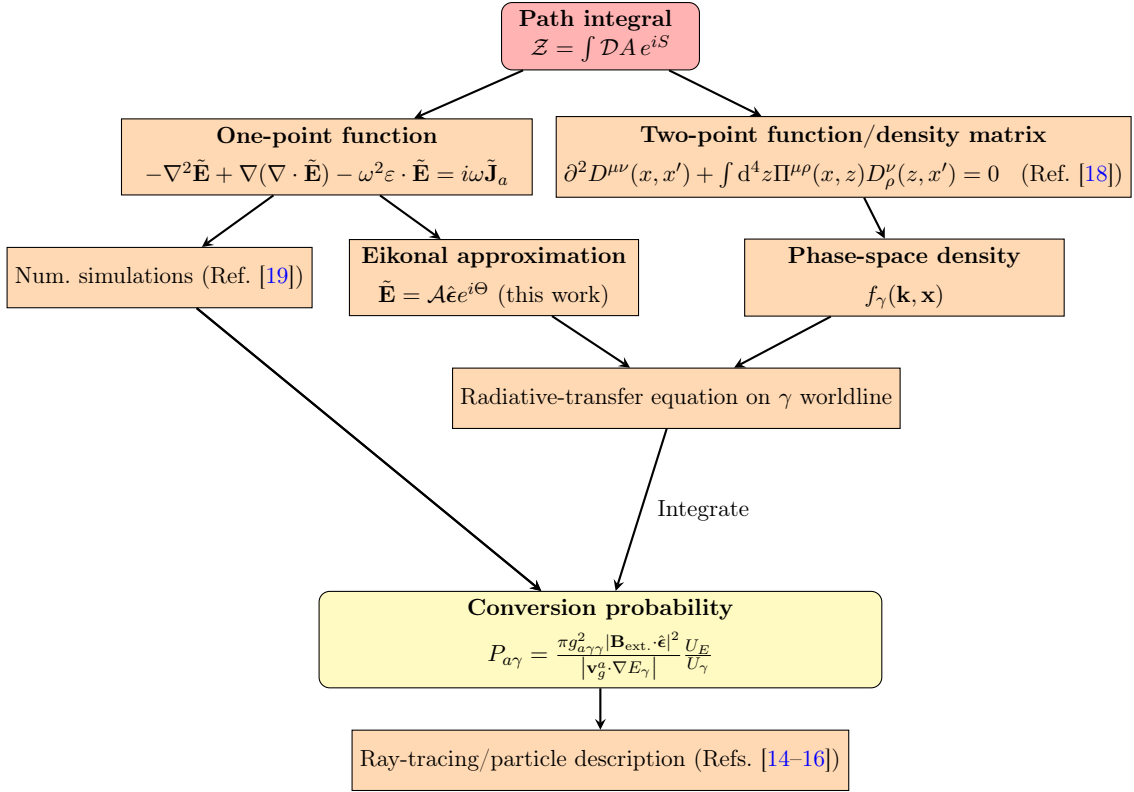
\begin{figure}[t!]
\centering
\hspace*{-1.5cm}
\normalsize
\tikzset{every picture/.style={scale=0.8, transform shape}}
\begin{tikzpicture}[node distance=2cm and 3cm]

\node (start) [startstop, xshift=-15cm] {
\centering
\parbox{3cm}{\textbf{Path integral}\\ 
\centering $\mathcal{Z} = \int \mathcal{D} A\, e^{i S}$
}
};

\node (process1A) [process, below of=start, xshift=-4.5cm] {
\parbox{6.5cm}{
\centering \textbf{One-point function}\\[1ex]
$-\nabla^2 \tilde{\textbf{E}} + \nabla(\nabla \cdot \tilde{\textbf{E}}) - \omega^2\varepsilon \cdot \tilde{\textbf{E}} = i\omega\tilde{\textbf{J}}_a$
}
};

\node (process1B) [process, below of=start, xshift=4cm] {
\parbox{9.2cm}{ \centering \textbf{Two-point function/density matrix} \\[1ex]
$\partial^2 D^{\mu \nu}(x,x') + \int {\rm d}^4 z \Pi^{\mu \rho}(x,z) D_\rho ^\nu(z,x') =0$ \hspace{0.1cm} (Ref.~\cite{McDonald:2023ohd})
}
};

\node (process2Aa) [process, below of=process1A, xshift=-2.75cm] { Num.\ simulations (Ref.~\cite{Gines:2024ekm})};

\node (process2Ab) [process, below of=process1A, xshift=2.75cm] { \parbox{4.5cm}{ \centering \textbf{Eikonal approximation} \\ [1ex]
$\tilde{\textbf{E}} = \mathcal{A} \hat{\boldsymbol{\epsilon}} e^{i \Theta}$ (this work) }
};

\node (process2B) [process, below of=process1B, xshift=1cm] {
\centering
\parbox{5cm}{\centering \textbf{Phase-space density}\\[1ex] 
\centering $f_\gamma (\textbf{k},\textbf{x})$}
};

\node (process4) [process, below of=process2Ab, xshift=3cm] { Radiative-transfer equation on $\gamma$ worldline};

\node (stop) [stop, below of=start, yshift=-8cm] {

\parbox{9cm}{
\centering { \textbf{Conversion probability} }\\ [1ex]
 \centering $ P_{a\gamma} =  \frac{\pi g^2_{a \gamma \gamma}  \left|\textbf{B}_{\rm ext.}\cdot \hat{\boldsymbol{ \epsilon}} \right|^2}{\left| \textbf{v}^a_g \cdot \nabla E_\gamma \right|} \frac{U_E}{U_{\gamma}}$
   }
}; 

\node (process5) [process, below of=stop, xshift=0cm] {Ray-tracing/particle description (Refs.~\cite{Battye:2021xvt,Witte:2021arp,McDonald:2023shx})};

\draw [arrow] (start) -- (process1A);
\draw [arrow] (start) -- (process1B);
\draw [arrow] (process1A) -- (process2Aa);
\draw [arrow] (process1A) -- (process2Ab);
\draw [arrow] (process1B) -- (process2B);
\draw [arrow] (process2Ab) -- (process4);
\draw [arrow] (process2B) -- (process4);
\draw [arrow] (process4) -- node[midway, right] {\ Integrate} (stop); 
\draw [arrow] (process2Aa) -- (stop);
\draw [arrow] (process2Aa) -- (stop);
\draw [arrow] (stop) -- (process5);

\end{tikzpicture}
\caption{Flowchart demonstrating three equivalent pathways to compute the conversion probability $P_{a \gamma}$. The first is to solve the equation for the one-point function, i.e., the field $\textbf{E}$. This can be done either directly via full numerical simulations of the second-order differential equation, as has recently been done in Ref.~\cite{Gines:2024ekm}, or via a WKB/eikonal Ansatz (this work). In the latter approach, one can derive a first-order transport equation \eqref{eq:TransportEq}, which can then be integrated along the photon worldline to obtain $P_{a \gamma}$. Alternatively, one can work at the level of two-point functions, introducing phase-space densities via a Wigner transform, as shown in Ref.~\cite{McDonald:2023ohd}. This also leads to a transport equation, which can again be projected onto the photon worldline and integrated to obtain the conversion probability. All three approaches give consistent results for the conversion probability, which differ from that presented in Ref.~\cite{Millar2021}.}
\label{fig:Pathways}
\end{figure}

In this short paper, we show how the same result can be derived directly from with the wave equations of axion-electrodynamics at the level of the Wentzel--Kramers--Brillouin (WKB) approximation, wherein the axion and photon fields are approximated as local plane waves. In Fig.~\ref{fig:Pathways}, we illustrate the three equivalent independent pathways (Refs.~\cite{McDonald:2023ohd} and \cite{Gines:2024ekm}, and this work), which can be used to compute the resonant conversion probability.

The structure of this paper proceeds as follows. In Section \ref{Sec:transport}, we lay out the equations for axion-electrodynamics and show, by inserting a local plane-wave Ansatz for the photon, how these lead to a transport equation for the photon plasma eigenmodes. In Section \ref{sec:Worldline}, we show how to solve the resulting transport equation along photon worldlines and, in Sec.~\ref{sec:Continuity}, we use this to  derive a continuity equation for the outgoing photon flux. We then define a conversion probability as the ratio of the axion to photon fluxes. Finally, our conclusions are given in Section \ref{sec:conclusions}.

\section{Photon transport equation}\label{Sec:transport}

Axions couple to photons via the following interaction:
\begin{equation}
    \mathcal{L}_{a \gamma\gamma} = - \frac{1}{4}g_{a\gamma\gamma}aF_{\mu\nu}\tilde{F}^{\mu\nu} \,,
\end{equation}
where $g_{a\gamma\gamma}$ is the coupling, $a$ is the pseudo-scalar axion, $F_{\mu\nu}$ is the usual electromagnetic field strength tensor, and $\tilde{F}^{\mu\nu}$ is its dual. 
When this interaction is appended to the Maxwell Lagrangian of electrodynamics, one obtains the following equations for axion-electrodynamics (in the absence of free charges and currents):
\begin{subequations}
\begin{align}
\nabla \cdot \textbf{D}& =  - g_{a \gamma \gamma} \textbf{B} \cdot \nabla  a\,, \label{Gauss}\\
\nabla \times \textbf{B} - \dot{\textbf{D}} &=  g_{a \gamma \gamma}\dot{a} \textbf{B} - g_{a \gamma \gamma} \textbf{E}\times \nabla a\,,\label{curlB}\\
\nabla \cdot \textbf{B} &=0\,, \\
\dot{\textbf{B} } + \nabla \times \textbf{E}& =0 \label{Bianchi2}\,.
\end{align}
\end{subequations}
Herein, $\mathbf{E}$ and $\mathbf{B}$ are the electric and magnetic fields, $\mathbf{D}$ is the displacement field, and we have assumed the magnetic permeability $\mu = 1$. From Eq.~\eqref{curlB}, we can identify an effective axion current
\begin{equation}\label{eq:Ja}
    \textbf{J}_{a} = g_{a \gamma \gamma}\dot{a} \textbf{B} - g_{a \gamma \gamma} \textbf{E}\times \nabla a  \,.
\end{equation}
Taking the time derivative of Eq.~\eqref{curlB} and using Eq.~\eqref{Bianchi2}, we arrive at
\begin{equation}\label{eq:WaveEq1}
- \nabla^2 \, \textbf{E} + \nabla (\nabla \cdot \textbf{E}) + \ddot{\textbf{D}} = -\dot{\textbf{J}}_a\,.
\end{equation}

Next, we assume a stationary background, which allows us to project into frequency space. In addition, we assume a simple constitutive relation
\begin{equation}
    \tilde{\textbf{D}}(\omega, \textbf{x}) = \varepsilon(\omega,\textbf{x}) \cdot \tilde{\textbf{E}}(\omega, \textbf{x})\,,
\end{equation}
where $\varepsilon$ is the $3\times 3$ permittivity tensor and tildes denote the Fourier transforms with respect to time. Putting this together in frequency space, Eq.~\eqref{eq:WaveEq1} yields
\begin{align}\label{eq:PhotonWaveEq}
- \nabla^2 \, \tilde{\textbf{E}} + \nabla (\nabla \cdot \tilde{\textbf{E}}) - \omega^2 \varepsilon(\omega,\textbf{x}) \cdot \tilde{\textbf{E}} &=  i \omega \tilde{\textbf{J}}_a\,.
\end{align}
This can be solved in the Born approximation, wherein we expand $\tilde{\textbf{J}}_a$ about background fields. In the case of a purely magnetic background field $\textbf{B}_{\rm ext.}$, we have 
\begin{equation}
\tilde{\textbf{J}}_a = - i \omega g_{a \gamma \gamma} \omega \tilde{a}(\textbf{x},\omega) \textbf{B}_{\rm ext.}\,.
\end{equation}
We emphasise that any other current, arising from other particle species, e.g., gravitons~\cite{Berlin:2021txa,McDonald:2024nxj} or $C\! P$-even scalars~\cite{McDonald:2023ohd} can readily be substituted without changing the arguments that follow. 

The aim now is to derive a photon transport equation that describes the evolution of the photon along its worldline, sourced by the axion field. We consider a plane-wave-like solution for the incoming axion field of the form
\begin{equation}
    \tilde{a} = a_0 e^{i \textbf{k}_a\cdot \textbf{x}}\,,
\end{equation}
where $\textbf{k}_a$ is the axion three-momentum and $a_0$ is the amplitude of the axion field. Next, we insert a WKB-like Ansatz for the electric field 
\begin{equation}
    \label{eq:EAnsatz}
   \tilde{ \textbf{E}} = \mathcal{A} \, \hat{\boldsymbol{ \epsilon} } \, e^{i \Theta(\textbf{x})}\,,
\end{equation}
where $\hat{\boldsymbol{ \epsilon} }$ is a unit polarisation vector, $\mathcal{A}$ is an amplitude and $\Theta$ is a phase. Inserting this into the equation of motion \eqref{eq:PhotonWaveEq} yields
\begin{align}\label{eq:WKBEquation}
&\mathcal{D}_{ij}  \left(\mathcal{A} \hat{\boldsymbol{ \epsilon}}_j  \right) +  i \frac{ \partial \mathcal{D}_{i j}}{\partial \textbf{k}_l} \nabla_l (\mathcal{A} \hat{\boldsymbol{ \epsilon}}_j )  + 
i \frac{\left(\mathcal{A}\hat{\boldsymbol{ \epsilon}}_j  \right)}{2}
\frac{ \partial \mathcal{D}_{i j}}{\partial \textbf{k}_l \partial \textbf{k}_{l'}} \nabla_l \textbf{k}_{l '}  + 
\frac{1}{2
}\frac{ \partial \mathcal{D}_{i j}}{\partial \textbf{k}_l \partial \textbf{k}_{l'}} \nabla_l \nabla_{l'} (\mathcal{A} \hat{\boldsymbol{\epsilon}}_j) \nonumber \\
& \qquad =  g_{a \gamma \gamma} \omega^2 a_0 [\textbf{B}_{\rm ext.}]_i e^{i (\textbf{k}_a\cdot \textbf{x} - \Theta)} \,.
\end{align}
Here, we have defined the effective wave operator 
\begin{equation}\label{eq:WaveOp}
    \mathcal{D}_{ i j} = - \left| \textbf{k} \right|^2 \delta_{ij} + \textbf{k}_i \textbf{k}_j + \omega^2 \varepsilon_{i j}(\omega, \textbf{x})
\end{equation}
in the local momentum space spanned by the effective three-momentum vector of the photon
\begin{equation}\label{eq:GradTheta1}
    \textbf{k}  \equiv \nabla \Theta\,,
\end{equation}
defined as the gradient of the phase.

Next, we seek solutions order-by-order in the gradients appearing on the left-hand side of Eq.~\eqref{eq:WKBEquation}. The terms at zeroth order in gradients impose that  $\hat{\boldsymbol{ \epsilon}}$ is a zero eigenvector of $\mathcal{D}_{ij}$, with eigenvalues $\mathcal{H}$, so that
\begin{equation}\label{eq:EigenEq}
    \mathcal{D}_{ij} \hat{\boldsymbol{ \epsilon}}_j = \mathcal{H} \hat{\boldsymbol{ \epsilon}}_i = 0 \,.
\end{equation}
It follows that the first term in Eq.~\eqref{eq:WKBEquation} therefore plays the role of a constraint equation (cf. Ref.~\cite{McDonald:2023ohd}). This leaves a kinetic-like transport equation for the amplitude, whose left-hand side is composed of the second and third terms on the left-hand side of Eq.~\eqref{eq:WKBEquation}, which is sourced by the axion terms on the right-hand side. The fourth term on the left-hand side is higher order in gradients and will be neglected hereafter.

Projecting with the polarisation vector from the left, the resulting transport equation~\eqref{eq:WKBEquation} becomes
\begin{align}
& \Big[ \hat{\boldsymbol{ \epsilon}}^* \cdot (\partial_\textbf{k} \mathcal{D} )\cdot \hat{\boldsymbol{ \epsilon}} \Big] \cdot \nabla \mathcal{A}  + \Big[ \hat{\boldsymbol{ \epsilon}}^* \cdot (\partial_\textbf{k} \mathcal{D} )\cdot \nabla \hat{\boldsymbol{ \epsilon}} \, + \,   
 \frac{1}{2}\hat{\boldsymbol{ \epsilon}}_i^* \left(\partial_{\textbf{k}_l}  \partial_{\textbf{k}_{l'}}  \mathcal{D}_{i j}  \right) \hat{\boldsymbol{ \epsilon}}_j\nabla_l \textbf{k}_{l '} 
\Big]\mathcal{A} \nonumber \\
& \qquad = g_{ a \gamma \gamma} \omega^2 a_0 \left(\textbf{B}_{\rm ext.}\cdot \hat{\boldsymbol{ \epsilon}^*} \right) e^{i (\textbf{k}_a\cdot\textbf{x} - \Theta)}\,,\label{eq:AnsatzProjection}
\end{align}
wherein we have absorbed an overall constant complex phase into the amplitude $\mathcal{A}$.
Since the polarisation vectors are on-shell, it is readily confirmed by differentiating Eq.~\eqref{eq:EigenEq} that 
\begin{equation}\label{eq:vg1}
    \frac{\hat{\boldsymbol{ \epsilon}}^* \cdot (\partial_\textbf{k} \mathcal{D} )\cdot \hat{\boldsymbol{ \epsilon}}}{\partial_\omega \mathcal{H}} = \frac{\partial_\textbf{k} \mathcal{H}}{\partial_\omega \mathcal{H}} \equiv \textbf{v}^\gamma_g\,,
\end{equation}
where $\textbf{v}^\gamma_g = \partial\omega/\partial\textbf{k}$ is the photon group velocity. The 
transport equation can therefore be recast in the form
\begin{equation}\label{eq:TransportEq}
\textbf{v}_g^\gamma \cdot \nabla \mathcal{A}+ \chi \mathcal{A}  =  g_{a \gamma \gamma} \omega a_0 \left(\textbf{B}_{\rm ext.}\cdot \hat{\boldsymbol{ \epsilon}}^* \right)\frac{U_E}{U_\gamma}e^{i (\textbf{k}_a\cdot\textbf{x} - \Theta)}\,,
\end{equation}
where we divided \eqref{eq:AnsatzProjection} through by $\partial_\omega \mathcal{H}$ and used the result \cite{McDonald:2023ohd} $\partial_\omega \mathcal{H} =  \omega U_E/U_\gamma$ 
in which
\begin{equation}
    U_\gamma =  \frac{1}{4}|\tilde{\textbf{B}}|^2  +  \frac{1}{4}\tilde{\textbf{E}}^*_i  \partial_{\omega} (\omega \epsilon_{ i j }) \tilde{\textbf{E}}_j\,, \qquad   U_E = \frac{1}{4}|\tilde{\textbf{E}}|^2
\end{equation}
are the (phase-averaged) total and electric energy of the photon, respectively. In addition, we have defined
\begin{equation}
    \label{eq:chi}
 \chi =  \frac{1}{\partial_\omega \mathcal{H}}\Big[ \hat{\boldsymbol{ \epsilon}}^* \cdot (\partial_\textbf{k} \mathcal{D} )\cdot \nabla \hat{\boldsymbol{ \epsilon}} \, + \, \frac{1}{2}\hat{\boldsymbol{ \epsilon}}_i^* \left(\partial_{\textbf{k}_l}  \partial_{\textbf{k}_{l'}}  \mathcal{D}_{i j}  \right) \hat{\boldsymbol{ \epsilon}}_j\nabla_l \textbf{k}_{l '}
\Big]\,.
\end{equation}

We now recognise the first term in Eq.~\eqref{eq:TransportEq} as the convective derivative along worldlines, whose tangent vector is given by the group velocity $\textbf{v}_g^\gamma$. In the object $\chi$, the second terms proportional to $\nabla_{l} \textbf{k}_{\ell'}$ correspond to the distortion of geodesic bundles, which experience so-called expansion, shear and twist (see, e.g., Refs.~\cite{Misner:1973prb} and \cite{McDonald:2014yfg}) and cause a change in the intensity of light as the ray bundle changes shape. These concepts are familiar from geometric optics in curved spacetime and Raychaudhuri's equation \cite{Hawking:1973uf}.

After a few short steps, detailed in Appendix~\ref{sec:AppContinuity}, we can also recast Eq.~\eqref{eq:AnsatzProjection} in the form of a continuity equation for the photon energy density $U_\gamma$:
\begin{equation}\label{eq:ContinuityUgamma}
\textbf{v}_g^\gamma \cdot \nabla U_\gamma +  \left(\nabla \cdot \textbf{v}_g^\gamma \right) U_\gamma  =  \frac{1}{4}g_{a \gamma \gamma} \omega^2 a_0 \left(\textbf{B}_{\rm ext.}\cdot \hat{\boldsymbol{ \epsilon}}^* \right) e^{i (\textbf{k}_a\cdot\textbf{x} - \Theta)} \mathcal{A}^* + \text{H.c.} \, ,
\end{equation}
where $``\text{H.c.}"$ denotes the addition of the hermitian conjugate of the right-hand side to itself. Note in particular the resemblance of the left-hand side of Eq.~\eqref{eq:ContinuityUgamma} to the Euler equation for a compressible fluid.

\section{Photon worldline method}\label{sec:Worldline}

Using a method of characteristics, we can follow the steps outlined in Ref.~\cite{McDonald:2023ohd} to solve the transport equation~\eqref{eq:TransportEq} along worldlines where $\mathcal{H}(\textbf{x}(s),\textbf{k}(s))  = 0$, corresponding to Hamilton's equations
\begin{equation}\label{eq:Hamilton}
    \frac{{\rm d}\textbf{x}}{{\rm d}s} = \textbf{v}_g^\gamma\,, \qquad \frac{{\rm d}\textbf{k}}{{\rm d}s} = - \nabla E_\gamma\, .
\end{equation}
Along these worldlines, Eq.~\eqref{eq:TransportEq} becomes
\begin{equation}
    \frac{{\rm d}\mathcal{A}}{{\rm d}s} + \chi \mathcal{A} = g_{a \gamma \gamma} \omega a_0 \left(\textbf{B}_{\rm ext.}\cdot \hat{\boldsymbol{ \epsilon}}^* \right) \frac{U_E}{U_\gamma} e^{i (\textbf{k}_a\cdot\textbf{x} - \Theta(s))},
\end{equation}
whose solution is
\begin{equation}
    \label{eq:Aint}
    \mathcal{A}(s) = \int^s_{ -\infty} {\rm d}s' \, e^{\int^{s'}_s {\rm d}s'' \chi(s'')}  g_{a \gamma \gamma} \omega a_0 \left(\textbf{B}_{\rm ext.}\cdot \hat{\boldsymbol{ \epsilon}}^* \right) \frac{U_E}{U_\gamma} e^{i (\textbf{k}_a\cdot\textbf{x}(s') - \Theta(s'))}\,,
\end{equation}
so that $\mathcal{A}(s) \rightarrow 0$ as $s \rightarrow - \infty$. This integral expression is, of course, also able to capture wave-optical interference effects between different resonances, as described in Ref.~\cite{Brahma:2023zcw}.  The result \eqref{eq:Aint} therefore augments the kinetic theory description \cite{McDonald:2023ohd}, which computed the conversion across a single level crossing. In Appendix~\ref{sec:AppHomogeneous}, we also verify that the integral \eqref{eq:Aint} reduces to the usual expression \cite{Raffelt:1987im} for axion-photon mixing in a \textit{homogeneous} background

\subsection{Stationary phase and resonant conversion}

The integral \eqref{eq:Aint} gives the most general expression for $\mathcal{A}$. However, we can proceed further by applying the method of stationary phase to the $s'$ integration. The phase is
\begin{equation}\label{eq:PhaseDef}
    \Psi(s) \equiv \textbf{k}_a\cdot\textbf{x}(s) - \Theta(\textbf{x}(s))\,,
\end{equation}
and the integral is then given, up to a constant overall phase, by 
\begin{equation}\label{eq:StatPhaseApprox}
    \mathcal{A}(s \rightarrow \infty) =  \, e^{ - \int^{\infty}_{s_c} {\rm d}s'' \chi(s'')}  \, g_{a \gamma \gamma} \omega a_0 \left(\textbf{B}_{\rm ext.}\cdot \hat{\boldsymbol{ \epsilon}}^* \right)\frac{U_E}{U_\gamma}\sqrt{\frac{2 \pi}{\left| \Psi''(s_c) \right|}}\,,
\end{equation}
wherein we we have integrated up to $s \rightarrow \infty$. Here, $s_c$ is the location of the stationary phase, which occurs at
\begin{equation}
    \label{eq:stat_phase_loc}
 \Psi'(s_c)    = \frac{{\rm d} \textbf{x}}{{\rm d}s} \cdot \left( \textbf{k}_a - \textbf{k}(s_c) \right) = 0\,,
\end{equation}
where we have used that $\Theta'(\textbf{x}(s)) = \textbf{x}'(s) \cdot \nabla\Theta  \equiv \textbf{x}'(s)\cdot \textbf{k}$ and the identity \eqref{eq:GradTheta1}.  Equation~\eqref{eq:stat_phase_loc} implies the stationary condition 
\begin{equation}
    \textbf{k}_a = \textbf{k}_\gamma\,,
\end{equation}
as expected.\footnote{There is another ``spurious'' stationary phase point when $\textbf{x}'(s)$ is perpendicular to the momentum splitting $\textbf{k}_a - \textbf{k}$, but this is unphysical and represents a breakdown of WKB. This is because the first term in Eq.~\eqref{eq:TransportEq} would then vanish, contrary to the fact that it is supposed to be dominant over the high-order terms on the left-hand side of Eq.~\eqref{eq:WKBEquation}.} The second derivative of the phase appearing in Eq.~\eqref{eq:StatPhaseApprox} is given by
\begin{equation}
    \Psi''(s) =  \frac{{\rm d}^2 \textbf{x}}{{\rm d}s^2} \cdot \left( \textbf{k}_a - \textbf{k}(s) \right)  -\frac{{\rm d} \textbf{x}}{{\rm d}s} \cdot \textbf{k}'(s)\,.
\end{equation}
Clearly, the first term vanishes on resonance when $\textbf{k}_a = \textbf{k}(s)$, and we can use Hamilton's equations \eqref{eq:Hamilton} to write $\textbf{k}'(s) = - \nabla E_\gamma$ and $\textbf{x}'(s) = \textbf{v}_g^\gamma$ to obtain
\begin{equation}
    \left| \Psi''(s_c) \right|  = \left| \textbf{v}_g^\gamma \cdot \nabla E_\gamma\right|.
\end{equation}
Putting this together, we arrive at the following expression for the squared amplitude:
\begin{equation}\label{eq:ESupress}
    \left| \mathcal{A}(s \rightarrow \infty)\right|^2 =  \, e^{ - 2 \int^{\infty}_{s_c} {\rm d}s' \text{Re}\,\chi(s')}  \, g^2_{a \gamma \gamma} \omega^2 a^2_0 \left| \textbf{B}_{\rm ext.}\cdot \hat{\boldsymbol{ \epsilon}} \right|^2 \frac{U_E^2}{U_\gamma^2}\frac{ 2\pi}{\left| \textbf{v}_g^\gamma \cdot \nabla E_\gamma \right|}\,.
\end{equation}
Having taken the limit $s\to\infty$ and performed the Gaussian integration that arises in the stationary phase approximation over the whole real line,  this result gives the asymptotic value of the electric field, which is suppressed by the exponential prefactor, corresponding to the reduction in the intensity of the electric field due to geodesic expansion, as discussed at the end of Sec.~\ref{Sec:transport}. However, as we will see in the next section, the total photon flux through an infinitesimal volume around the photon worldline remains conserved and is insensitive to these effects. We can also evaluate the integral \eqref{eq:Aint} on the half-real line up to $s_c^+$, which gives an additional factor of $1/2$ and no exponential prefactor, i.e.,
\begin{equation}\label{eq:StatPhaseHalf}
  \mathcal{A}(s_c^+) =  \, \frac{1}{2}  \, g_{a \gamma \gamma} \omega a_0 \left( \textbf{B}_{\rm ext.}\cdot \hat{\boldsymbol{ \epsilon}}^* \right) \frac{U_E}{U_\gamma} \sqrt{  \frac{ 2 \pi}{\left| \textbf{v}_g^\gamma \cdot \nabla E_\gamma \right|}}\,.
\end{equation}
We will make use of the result \eqref{eq:StatPhaseHalf} in the next section.

\section{Continuity equation and conversion probability}\label{sec:Continuity}

In this section, we develop the continuity equation for the photon flux, which allows us to define a probability for axion-photon conversion. We begin with the Poynting flux for the photon, which we identify as 
\begin{equation}
    \textbf{S}_\gamma = \textbf{v}_g^\gamma U_\gamma \, .
\end{equation}
This allows us to rewrite the continuity equation \eqref{eq:ContinuityUgamma} as
\begin{equation}
\label{eq:divS}
\nabla \cdot  \textbf{S}_\gamma  = \frac{1}{4} g_{a \gamma \gamma} \omega a_0 \left(\textbf{B}_{\rm ext.}\cdot \hat{\boldsymbol{ \epsilon}}^* \right) e^{i (\textbf{k}_a\cdot\textbf{x} - \Theta)} \mathcal{A}^* + \text{H.c.} \, .
\end{equation}

It will also be useful to define the axion/photon group velocities $\textbf{v}_g^{\gamma/ a}$, respectively. It was shown in  Ref.~\cite{McDonald:2023ohd} that the conversion surface has normal parallel to $\nabla E_\gamma$, since it is defined by the level surfaces of  $E_a(\textbf{k}) - E_\gamma(\textbf{x},\textbf{k})$. Hence, we can write 
\begin{equation}\label{eq:Angles}
    {\rm d}\boldsymbol{\Sigma}_\textbf{k} \cdot \textbf{v}^{\gamma/a}_g = \cos \theta_{\gamma/a} {\rm d}\Sigma_\textbf{k} v^{\gamma/a}_g\,,
\end{equation}
where $\theta_{\gamma/a}$ is the angle between the photon/axion group velocity and $\nabla E_\gamma$, which is normal to the resonant conversion surface element ${\rm d} \Sigma_\textbf{k}$.

Integrating the divergence of the Poynting flux over an enclosing volume $\mathcal{V}$ with bounding surface element ${\rm d} \textbf{A}$, Eq.~\eqref{eq:divS} yields 
\begin{equation}\label{eq:continuity3}
\int {\rm d} \textbf{A}\cdot \textbf{S}_\gamma  = \int {\rm d} \mathcal{V} \left[ \frac{1}{4} g_{a \gamma \gamma} \omega a_0 \left(\textbf{B}_{\rm ext.}\cdot \hat{\boldsymbol{ \epsilon}}^* \right) e^{i (\textbf{k}_a\cdot\textbf{x} - \Theta)}  \mathcal{A}^*  + \text{H.c.} \,\right].
\end{equation}
The integral will be dominated by stationary phase, so we can expand the right-hand volume integral about the critical surface by writing
\begin{equation}
   \int {\rm d} \mathcal{V} = \int {\rm d}\Sigma_\textbf{k} \, \int_{-\infty}^\infty {\rm d} s \, v^\gamma_g \cos \theta_\gamma \, ,
\end{equation}
where we recall that $\theta_\gamma$ is the angle between the photon group velocity and ${\rm d}\Sigma_\textbf{k}$. Next, we can perform the ${\rm d}s$ integral using the method of stationary phase and following steps similar to those carried out in Eqs.~\eqref{eq:PhaseDef}--\eqref{eq:StatPhaseHalf}. This leaves (again up to overall phases that will cancel with the contributions from $\mathcal{A}^*$)
\begin{equation}
\int {\rm d} \textbf{A}\cdot \textbf{S}_\gamma  = \int {\rm d} \Sigma_\textbf{k} v^\gamma_g \cos \theta_\gamma  \left[  \frac{1}{4} g_{a \gamma \gamma} \omega a_0 \left(\textbf{B}_{\rm ext.}\cdot \hat{\boldsymbol{ \epsilon}}^* \right)
\sqrt{\frac{2 \pi}{\left| \textbf{v}_g^\gamma \cdot \nabla E_\gamma\right|}}
 \mathcal{A}^*(s_c^+) + \text{H.c.} \, \right]\, , 
\end{equation}
where all quantities in the integral are evaluated on the critical surface. Next, we insert the expression \eqref{eq:StatPhaseHalf} for $\mathcal{A}(s_c^+)$ on the critical surface, cancel the factors of $v^\gamma_g \cos \theta_\gamma $ between the numerator and the denominator, and add the Hermitian conjugate to arrive at 
\begin{equation}
\int {\rm d} \textbf{A}\cdot \textbf{S}_\gamma  = \int {\rm d} \Sigma_\textbf{k}  \left[  \frac{1}{2} \omega^2 a_0^2 \frac{\pi g^2_{a \gamma \gamma}   \left| \textbf{B}_{\rm ext.}\cdot \hat{\boldsymbol{ \epsilon}}\right|^2 }{\left|  \nabla E_\gamma\right|}
\frac{U_E}{U_\gamma}  \, \right]\, . 
\end{equation}
Finally, we insert a factor of $v^a_g\cos \theta^a $ in the numerator and the denominator, and use the definition of the axion flux density 
\begin{equation}
    \textbf{S}_a  = \textbf{v}_g^a \, U_a = \frac{1}{2} \textbf{v}_g^a  \omega^2 a_0^2 \, ,
\end{equation} 
to write 
\begin{equation}
\label{eq:SdotP_int}
\int {\rm d} \textbf{A}\cdot \textbf{S}_\gamma  = \int {\rm d} \Sigma_\textbf{k} \cdot \textbf{S}_a P_{a \gamma}\, ,  
\end{equation}
where we have \textit{defined}
\begin{equation}
   P_{a\gamma} =  \frac{ \pi g^2_{a \gamma \gamma}  \left|\textbf{B}_{\rm ext.}\cdot \hat{\boldsymbol{ \epsilon}} \right|^2}{\left| \textbf{v}^a_g \cdot \nabla E_\gamma \right|} \frac{U_E}{U_\gamma}\,.
\label{eq:prob}
\end{equation}
Herein, we have used the result \eqref{eq:Angles} to reformulate the denominator in terms of the dot product between the axion phase velocity $\textbf{v}^a_g \equiv \textbf{k}/\omega$ and $\nabla E_\gamma$, rather than the dot product with the \textit{photon} group velocity. This is precisely the expression for the conversion probability derived in Ref.~\cite{McDonald:2023ohd} by means of kinetic theory, but reproduced here from the wave equations of axion-electrodynamics directly.

\begin{figure}
    \centering
    \includegraphics[width=0.5\linewidth]{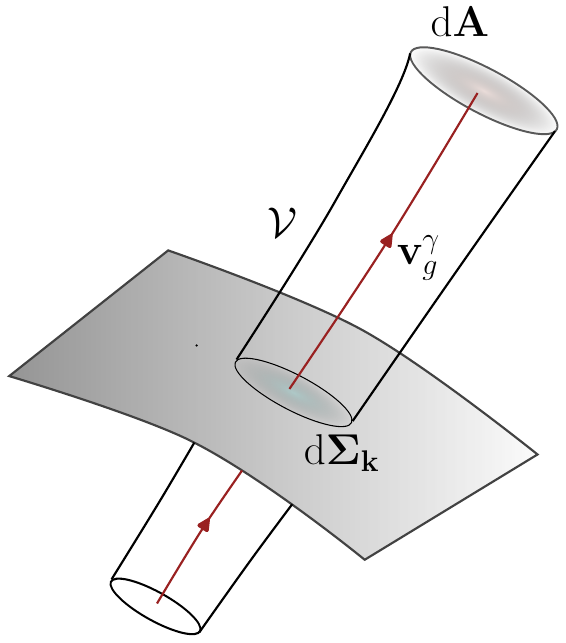}
    \caption{A tubular integration volume $\mathcal{V}$ around the photon worldline with tangent $\textbf{v}_g^\gamma$. The end cap gives the surface ${\rm d}\textbf{A}$ through which the photon flux $\textbf{S}_\gamma$ flows, while the axion flux $\textbf{S}_a$ passes through the critical surface element ${\rm d }\boldsymbol{\Sigma}_\textbf{k}$.  }
    \label{fig:tubular}
\end{figure}

We see from Eqs.~\eqref{eq:SdotP_int} and \eqref{eq:prob} that the geodesic expansion does not affect the integrated photon flux, as expected by conservation. Indeed, by drawing an infinitesimal tubular region $\mathcal{V}$ around the photon worldline (as illustrated in Fig.~\ref{fig:tubular}), we see that
\begin{equation}
    {\rm d}\textbf{A} \cdot \textbf{S}_\gamma = {\rm d}\boldsymbol{\Sigma}_\textbf{k} \cdot \textbf{S}_a P_{a \gamma} \, , 
\end{equation}
which is independent of the choice of ${\rm d} \textbf{A}$, as required by flux conservation. This is equivalent to Liouville's theorem discussed in Ref.~\cite{McDonald:2023shx}, i.e., the phase-space density of photons should be conserved along rays. Hence, although the rays diverge, causing a decrease in the amplitude of $\textbf{E}$ [see Eq.~\eqref{eq:ESupress}], the group velocity grows such that the $\textbf{S}_\gamma$ flowing through a surface remains constant. In other words, the spatial spread of photons is compensated by the bunching in velocity/momentum space of photons around $v_g^\gamma = 1$ at infinity, conserving the infinitesimal phase-space volume. 

\subsection{Application to strongly magnetised plasmas}

To apply the expression for the conversion probability~\eqref{eq:prob} to a particular medium, we require the relevant polarisation eigenvectors, the energy gradient and the relative electric power $U_E/U$, all of which are determined straightforwardly  from the photon dispersion relations and polarisation vectors in the medium in question. For example, for the Langmuir LO mode in a strongly magnetised plasma \cite{Witte:2021arp,McDonald:2023shx,McDonald:2023ohd}, one has
\begin{align}
    &U_E^{\rm LO}/U_\gamma^{\rm LO}  = 1/2 \,, \\ 
    \nonumber \\
    &E_{\rm LO}^2  = \frac{1}{2} \left( |\textbf{k}|^2 + \omega_p^2 + \sqrt{|\textbf{k}|^4  + \omega_p^4 + 2\omega_p^2 |\textbf{k}|^2(1 - 2 \cos^2 \theta)} \right)\,, \\
    \nonumber \\
   &\hat{\boldsymbol{\epsilon}}_{\rm LO} =\frac{-1}{\sqrt{1+\frac{\omega_p^4\cos^2\theta\sin^2\theta}{(\omega^2-\omega_p^2\cos^2\theta)^2}}}\left({\bf \hat y}-\frac{\omega_p^2\cos\theta\sin\theta}{\omega^2-\omega_p^2\cos^2\theta}{\bf \hat z}\right ) \,. \label{eq:EpsilonLO}
\end{align}
These have been expressed in coordinates where, without loss of generality, we have $\textbf{k}=(0,0,k)$ and $\textbf{B}_{\rm ext.}$ lies in the $y$-$z$ plane, with $\textbf{B}_{\rm ext.} = \left| \textbf{B}_{\rm ext.} \right|( - \sin \theta \hat{\textbf{y}} +  \cos \theta \hat{\textbf{z}})$, so that $\theta$ gives the angle between the photon momentum and the magnetic field. This leads to
\begin{equation}
P^{\rm LO}_{a \gamma } = \frac{\pi}{2} \frac{ g_{a \gamma \gamma}^2\left| \textbf{B}_{\rm ext.} \right|^2E_\gamma ^4 \sin ^2 \theta }{\cos ^2 \theta
   \, \omega _p^2 \left(\omega _p^2-2 E_\gamma ^2\right)+E_\gamma ^4} 	\frac{1}{\left| \textbf{v}^a_g \cdot \nabla E_\gamma  \right|} \,,
\end{equation}
reproducing the expression in Ref.~\cite{McDonald:2023ohd}.

\section{Conclusions}\label{sec:conclusions}

In this paper, we have demonstrated how the conversion probability for resonant axion-photon mixing can be obtained from the wave equations of classical axion-electrodynamics. The key step is to derive a transport equation projected along the photon worldline, which can be solved via a method of stationary phase. This reproduces the same result carried out in kinetic theory in Ref.~\cite{McDonald:2023ohd}, which was further supported by numerical simulations in Ref~\cite{Gines:2024ekm}. Taken together, these provide three independent checks of the validity of the axion-photon conversion formula~\eqref{eq:prob}. Note that these three results agree with one another but disagree with Ref.~\cite{Millar2021}. In addition, the results presented here also allow for the integrated effects of mixing along the photon worldline away from the resonance, which would have relevance for the interference effects between nearby level crossings considered in Ref.~\cite{Brahma:2023zcw}.

This formalism, as with our previous work \cite{McDonald:2023ohd},  is straightforwardly adaptable to mixing of any light particles by computing the  effective source current [see Eq.~\eqref{eq:Ja}] for the particle species in question.  

In summary, we have provided further evidence that the computation of resonant axion-photon mixing is under good control, allowing for the focus now to shift to the remaining uncertainties in the study of axion-photon mixing in neutron star magnetospheres, which are largely astrophysical, i.e., the structure of the magnetosphere itself.

\acknowledgments

We thank Sam Witte for many useful discussions, comments on the draft and advanced sight of the manuscript Ref.~\cite{Gines:2024ekm}. We thank Bj{\"o}rn Garbrecht for collaboration during the early phases of this project. JIM is grateful for discussions at the Barolo Astroparticle Meeting in June 2024, organised with support from the University of Torino and INFN. JIM acknowledges support form the Science and Technology
Facilities Council (STFC) [Grant No. ST/X00077X/1]. He also thanks the University of Oxford for hospitality. PM is supported by a United Kingdom Research and Innovation (UKRI) Future Leaders Fellowship [Grant No.\ MR/V021974/2] and the Science and Technology
Facilities Council (STFC) [Grant No. ST/X00077X/1].

\appendix

\section{Continuity Equation}\label{sec:AppContinuity}

To derive the transport equation \eqref{eq:ContinuityUgamma}, we begin by multiplying Eq.~\eqref{eq:AnsatzProjection} from the left by $\mathcal{A}^*$ and then add the Hermitian conjugate to arrive at
\begin{equation}\label{eq:ASquaredAppendix}
(\hat{\boldsymbol{ \epsilon}}^* \cdot (\partial_\textbf{k} \mathcal{D} )\cdot \hat{\boldsymbol{ \epsilon}}) \cdot \nabla |\mathcal{A}|^2+ 2 \text{Re}(\bar{\chi} ) \left| \mathcal{A} \right|^2  =  g_{a \gamma \gamma} \omega^2 a_0 \left(\textbf{B}_{\rm ext.}\cdot \hat{\boldsymbol{ \epsilon}}^* \right) e^{i (\textbf{k}_a\cdot\textbf{x} - \Theta)}\, \mathcal{A}^* + \text{H.c.} \, . 
\end{equation}
where $\bar{\chi}\equiv \chi\partial_\omega \mathcal{H}$, i.e.,
\begin{equation}
    \bar{\chi} = \hat{\boldsymbol{ \epsilon}}^* \cdot (\partial_\textbf{k} \mathcal{D} )\cdot \nabla \hat{\boldsymbol{ \epsilon}} \, + \, \frac{1}{2}\hat{\boldsymbol{ \epsilon}}_i^* \left(\partial_{\textbf{k}_l}  \partial_{\textbf{k}_{l'}}  \mathcal{D}_{i j}  \right) \hat{\boldsymbol{ \epsilon}}_j\nabla_l \textbf{k}_{l '}\, .
\end{equation}
Next, we notice that 
\begin{align}
    \nabla \cdot (\hat{\boldsymbol{ \epsilon}}^* \cdot (\partial_\textbf{k} \mathcal{D} )\cdot \hat{\boldsymbol{ \epsilon}} )  &=  \hat{\boldsymbol{ \epsilon}}^* \cdot (\partial_\textbf{k} \mathcal{D} )\cdot \nabla \hat{\boldsymbol{ \epsilon}} \, + \nabla\hat{\boldsymbol{ \epsilon}}^* \cdot (\partial_\textbf{k} \mathcal{D} )\cdot  \hat{\boldsymbol{ \epsilon}} \, + \, \hat{\boldsymbol{ \epsilon}}_i^* \left(\partial_{\textbf{k}_l}  \partial_{\textbf{k}_{l'}}  \mathcal{D}_{i j}  \right) \hat{\boldsymbol{ \epsilon}}_j\nabla_l \textbf{k}_{l '} 
 \nonumber \\
& \equiv  2 \text{Re}(\bar{\chi} ) \, , 
\end{align}
where, in deriving the third term on the right-hand side, we used the chain rule $\nabla_\ell \partial_{\textbf{k}_\ell} \mathcal{D}_{ij} =\left(\partial_{\textbf{k}_l}  \partial_{\textbf{k}_{l'}}  \mathcal{D}_{i j}  \right) \nabla_l \textbf{k}_{l '} $.  From this, we can immediately apply the product rule to the left hand side of \eqref{eq:ASquaredAppendix} and write 
\begin{equation}\label{eq:DivEquation}
\nabla\cdot \left( (\hat{\boldsymbol{ \epsilon}}^* \cdot (\partial_\textbf{k} \mathcal{D} )\cdot \hat{\boldsymbol{ \epsilon}})  \left| \mathcal{A}\right|^2\right) =  g_{a \gamma \gamma} \omega^2 a_0 \left(\textbf{B}_{\rm ext.}\cdot \hat{\boldsymbol{ \epsilon}}^* \right) e^{i (\textbf{k}_a\cdot\textbf{x} - \Theta)}\, \mathcal{A}^* + \text{H.c.} \, . 
\end{equation}
We now make use of the following identities, namely the result \eqref{eq:vg1}, the definition of the photon-electric energy 
\begin{equation}
    U_E = \frac{1}{4}|\tilde{\textbf{E}}|^2  = \frac{1}{4}|\mathcal{A}|^2 \, , 
\end{equation}
and the identity
\begin{equation}
   \partial_\omega \mathcal{H} =  \frac{U_\gamma}{U_E} \omega\,,
\end{equation}
as derived in our previous work \cite{McDonald:2023ohd}, to write 
\begin{align}\label{eq:StoredEnergyASquared}
    (\hat{\boldsymbol{ \epsilon}}^* \cdot (\partial_\textbf{k} \mathcal{D} )\cdot \hat{\boldsymbol{ \epsilon}})  \left| \mathcal{A}\right|^2 & =  \frac{ \hat{\boldsymbol{ \epsilon}}^* \cdot (\partial_\textbf{k} \mathcal{D}) \cdot \hat{\boldsymbol{ \epsilon}} }{\partial_\omega \mathcal{H}}  \partial_\omega \mathcal{H}  \left| \mathcal{A}\right|^2 \nonumber \\
    & = \omega \textbf{v}_g^\gamma  \frac{U_\gamma}{U_E} \left| \mathcal{A}\right|^2 \nonumber \\
     &   = 4 \omega \textbf{v}_g^\gamma  U_\gamma  .
\end{align}
Inserting \eqref{eq:StoredEnergyASquared} into \eqref{eq:DivEquation}, one arrives immediately at  
\begin{equation}
\nabla\cdot \left(\textbf{v}_g^\gamma U_\gamma \right) =  \frac{1}{4} g_{a \gamma \gamma} \omega a_0 \left(\textbf{B}_{\rm ext.}\cdot \hat{\boldsymbol{ \epsilon}}^* \right) e^{i (\textbf{k}_a\cdot\textbf{x} - \Theta)}\, \mathcal{A}^* + \text{H.c.} \, . 
\end{equation}
Finally, expanding the left-hand side using the product rule gives precisely Eq.~\eqref{eq:ContinuityUgamma}.

\section{Relativistic mixing in homogeneous backgrounds}\label{sec:AppHomogeneous}

In homogeneous backgrounds, we have $\chi =0$, and particles propagate in straight lines between $s = 0$ and $s = \ell$, say. The integral can then be evaluated straightforwardly between these limits to give
\begin{equation}
    |\mathcal{A}|^2 = \frac{g^2_{a \gamma \gamma } \omega^2 a_0^2 B^2_\parallel}{(k_a - k_\gamma)^2} \sin^2( (k_a - k_\gamma) \ell/2 )\,,
\end{equation}
where $B_{\parallel}$ is the component of $\mathbf{B}$ parallel to the polarisation vector $\hat{\boldsymbol{ \epsilon}}$.

In the relativistic limit and in a dilute plasma, one has $k_{a} = \omega - m_a^2/(2 \omega) $ and $k_\gamma = \omega - \omega_p^2/(2\omega)$ so that $k_a - k_\gamma =  \omega_p^2/(2\omega) -m_a^2/(2\omega) \equiv \ell^{-1}_{\rm osc.}$, which we identify as the usual (inverse) oscillation length. This allows us to write 
\begin{equation}
    P^{\rm hom.}_{a \gamma}(\ell) = \left( \frac{g_{a\gamma \gamma} B_{\parallel}}{\ell_{\rm osc.}^{-1}} \right)^2 \sin^2\left(\frac{\ell}{2\ell_{\rm osc.}}\right),
\end{equation}
reproducing the standard formula for axion-photon oscillation~\cite{Raffelt:1987im} in a homogeneous background in the linearised coupling limit and providing a useful consistency check of our results.

\bibliography{bibliography.bib}{}
\bibliographystyle{JHEP}

\end{document}